\documentclass{cargese}\usepackage{graphicx}
\let\footnote\savefootnote
\let\footnotetext\savefootnotetext 
\let\footnoterule\savefootnoterule 
\normallatexbib

\def\xxx#1           {hep-th/#1}
\def\npb#1(#2)#3     {Nucl. Phys. {\bf B#1} (#2) #3 }
\def\plb#1(#2)#3     {Phys. Lett. {\bf #1B} (#2) #3 }
\def\prl#1(#2)#3     {Phys. Rev. Lett.{\bf #1} (#2) #3 }
\def\cqg#1(#2)#3     {Class. Quant. Grav. {\bf #1} (#2) #3 }
\def\jhep#1(#2)#3    {JHEP {\bf #1} (#2) #3 }

\def\exp{\mbox{exp}}
\def\RR{R$\otimes$R }   
\def\NSNS{NS$\otimes$NS }
\def\w    {\,_\wedge}
\def\be              {\begin{equation}}
\def\ee              {\end{equation}}
\def\ba              {\begin{eqnarray}}
\def\ea              {\end{eqnarray}}
\def\tr{{\rm tr}}

\begin{document}
\articletitle[]{WZ Couplings of D-branes and O-planes}
\author{B. Stefa\'nski, jr.}
\affil{Department of Applied Mathematics and Theoretical Physics\\
Cambridge University, Silver Street, Cambridge CB3 9EW, UK }
\email{B.Stefanski@damtp.cam.ac.uk}

 \begin{abstract}
In this short note we review the construction and role of Wess-Zumino
 couplings of Dirichlet branes and Orientifold planes, and show how
these combine to give the Green-Schwarz anomaly cancelling terms.
 \end{abstract}

The low energy effective theory of a D-brane has two types of couplings, the
Dirac-Born-Infeld (DBI) and Wess-Zumino (WZ) terms. The former
describes the coupling of the vector potential and scalars coupling to
the \NSNS fields, while the latter gives the coupling of the vector
potential and the pullback of the curvature fields to the \RR
potentials. The first D-brane WZ term~\cite{PolRR} simply states that
a D$p$-brane is charged under $C^{(p+1)}$, the $p+1$-form \RR potential
\be
S_{WZ}^D=\mu_p\int_{{\cal B}_p}C^{(p+1)}\,,
\ee
where $\mu_p$ is the charge density of a D-brane and ${\cal B}_p$ its
worldvolume. It was then shown that D-branes couple to lower dimensional \RR
potentials~\cite{Lsd} and that a `brane within brane'
picture emerged
\be
S_{WZ}^D=\mu_p\int_{{\cal B}_p}C\w \tr\,\exp(i{\cal F}/2\pi)\,.
\ee
In the above $C$ represents a formal sum of \RR potentials, 
${\cal F}=F-B$, $F$ is the worldvolume gauge field and
$B$ the \NSNS two form.\footnote{${\cal F}$, rather than $F$, is
present in the action as the latter quantity is not gauge invariant.}

An $R^4$ term found in~\cite{VW} required, as a consequence of
duality, the presence of a gravitational WZ coupling~\cite{ber}.
Later the entire WZ action was found using the anomaly 
inflow mechanism~\cite{GHM,CY}
and reads
\be
S_{WZ}^D=\mu_p\int_{{\cal B}_p}C\w \tr\,\exp(i{\cal
F}/2\pi)\w\sqrt{\frac{\hat{A}(R_T)}{\hat{A}(R_N)}}\,,
\ee
where $\hat{A}$ is the Dirac or A-roof genus and $R_T,\;R_N$ are the
pull-backs of the tangent and normal bundle curvatures to the D-brane
world-volume, respectively. 

This coupling has a natural interpretation
within K-theory~\cite{MM}. In K-theory there is a natural
bilinear pairing of bundles given by the index of the Dirac operator on the
tensor product of the two bundles. For $E$ a bundle over a manifold
$X$, with $TX$ the tangent space of $X$, the map
\be
E\rightarrow \mbox{ch}(E)\sqrt{\hat{A}(TX)}\,,
\ee
is an isometry with respect to this pairing and the DeRham pairing 
in $H^*(X)$, the cohomology of $X$. In fact,following~\cite{Sen}, 
it was shown~\cite{wk} that D-brane charges are indeed classified by K-theory. 

The D-brane WZ couplings were confirmed by string amplitude
calculations. In~\cite{Craps} the four-form couplings
were computed at tree level. A one-loop amplitude in~\cite{MSS} and
tree-level amplitude in~\cite{bs} confirmed the presence of all the couplings,
while in~\cite{cr2} extra, non-anomalous couplings, as
well as the normal bundle contributions were determined.

In~\cite{dasgupta}, it was first observed that O-planes too
carry gravitational WZ couplings. For consider Type I theory, in which
16 D9-branes, their images and an O9-plane, fill the spacetime. The WZ
coupling of the theory is the
Green-Schwarz (GS) coupling~\cite{gs}. This differs from the
WZ couplings of 32 D9-branes, indicating that
O9-planes too have WZ couplings. By studying 
one-loop~\cite{MSS}, and tree-level~\cite{bs}
scattering amplitudes it was found that the WZ coupling of
Op-planes is
\be
S^O_{WZ}=-2^{p-4}\mu_p\int_{{\cal B}_p}C\w\sqrt{\frac{L(R_T/4)}{L(R_N/4)}}\,,
\ee
where $L$ is the Hirzebruch polynomial. To see that the O9-plane and
D9-brane WZ couplings match the GS term consider the following.
The massless chiral fields of Type I string theory are a neutral gravitino, a
neutral fermion of opposite chirality and SO(32) fermions. 
The total anomaly for this theory follows by descent 
from\footnote{Note that all of these are real fields.}
\ba
I&=&\left.\frac{1}{2}2\pi\left(\hat{A}(R)(\tr e^{iR/2\pi}-2)+\hat{A}(R)\tr
e^{iF/2\pi}\right)\right|_{\mbox{12-form}} \nonumber \\
&=&\pi\frac{1}{(4\pi)^2}(\tr R^2-\tr F^2)\w\; X_8\,,
\ea
where
\be
X_8=\frac{1}{(4\pi)^4}\left(\frac{2}{3}\tr F^4
+\frac{1}{12}\tr R^4+\frac{1}{48}(\tr R^2)^2-\frac{1}{12}\tr R^2\tr
F^2\right)\,.
\ee
In the units of~\cite{pv2} the action extracted from 
the string theory amplitudes~\cite{MSS,bs} is
\ba
S&=&
-\frac{1}{4\kappa_{10}^2}\int \mbox{d}C^{(2)}\!\w *\mbox{d}C^{(2)}
\nonumber \\& &
+\;\;\mu_9\int\left( \frac{2}{(4\pi)^2}C^{(6)}\!\w (\tr R^2-\tr F^2)
+C^{(2)}\!\!\w\, X_8\right)\,.\label{wzact}
\ea
Since $H=\mbox{d}C^{(2)}+\dots$ is gauge invariant and
$\mbox{d}C^{(6)}=*\,\mbox{d}C^{(2)}$ the gauge
transformation for $C^{(2)}$ is
\be
\delta C^{(2)}=4\mu_9\kappa_{10}^2(\omega_{2,Y}^1-\omega_{2,L}^1)\,,
\ee
hence~(\ref{wzact}) has an anomalous variation which
follows by descent from
\be
I_{WZ}=4\mu_9^2\kappa^2_{10}(\tr F^2-\tr R^2)\w X_8\,.
\ee
The Type I charge density satisfies
\be
(\mu_9\kappa_{10})^2=\frac{\pi}{2}\,,
\ee
and hence $I_{WZ}=I$ as required.

O-planes cannot couple to gauge fields so their WZ
couplings have to be purely gravitational. Hence it is a
consistency check for the D9-brane and GS gauge and mixed 
couplings to agree.


\begin{acknowledgments}
The author would like to thank the conference organisers for
financial support and a stimulating programme.
\end{acknowledgments}



%


\begin{chapthebibliography}{99}

\bibitem{PolRR} J. Polchinski, {\em Dirichlet-Branes and Ramond-Ramond
Charges}, \prl75(1995)4724,  \xxx9510017;

\bibitem{Lsd} M. Li, {\em Boundary States of D-branes and Dy-Strings}, 
\npb460(1996)351, hep-th/9510161; C. Schmidhuber, 
{\em D-Brane Actions}, \npb467(1996)146, hep-th/9601003; M.R. Douglas, 
{\em Branes within Branes}, \xxx9512077;

\bibitem{VW} C. Vafa and E. Witten, {\em A One Loop Test of String
Dualities}, \npb447(1995)261, \xxx9505053;

\bibitem{ber} M. Bershadsky, C. Vafa, V. Sadov, {\em D-Branes and
Topological Field Theories}, \npb463(1996)420, \xxx9511222;

\bibitem{GHM} M.B. Green, J.A. Harvey and G. Moore, {\em I-Brane Inflow
and Anomalous Couplings on D-branes}, \cqg14(1997)47, \xxx9605033;

\bibitem{CY} Y.-K.E. Cheung and Z. Yin, {\em Anomalies, Branes, and
Currents}, \npb517(1998)69, \xxx9710206;

\bibitem{MM} R. Minasian and G. Moore, {\em K-theory and Ramond-Ramond
Charge}, \jhep9711(1997)002, \xxx9710230;

\bibitem{dasgupta} K. Dasgupta, D.P. Jatkar, S. Mukhi, {\em Gravitational
Couplings and $Z_2$ Orientifolds}, \npb523(1998)465, \xxx9707224; 

\bibitem{Craps} B. Craps and F. Roose, {\em Anomalous D-brane and
Orientifold Couplings from the Boundary State}, \plb445(1998)150, \xxx9808074;

\bibitem{MSS} J.F. Morales, C.A. Scrucca, M. Serone, {\em Anomalous
Couplings for D-branes and O-planes}, \npb552(1999)291, \xxx9812071;

\bibitem{bs} B. Stefa\'nski, jr., {\em Gravitational Couplings of
D-branes and O-planes}, \npb548(1999)275, \xxx9812088;

\bibitem{cr2}  B. Craps and F. Roose, {\em (Non-)Anomalous D-brane and
O-plane couplings: the Normal Bundle}, \plb450(1999)358, \xxx9812149;

\bibitem{Sen} For a review see A. Sen, {\em Non-BPS States and Branes
in String Theory}, \xxx9904207;

\bibitem{wk} E. Witten, {\em D-Branes And K-Theory},
\jhep9812(1998)025, \xxx9810188;

\bibitem{gs} M.B. Green and J.H. Schwarz, {\em Anomaly Cancellations in
Supersymmetric D=10 Gauge Theory and Superstring Theory},
\plb149(1984)117;

\bibitem{pv2} J. Polchinski, {\em String Theory, Volume 2, Superstring
Theory and Beyond}, CUP 1998;




\end{chapthebibliography}
\end{document}